\documentclass[twocolumn,english,aps,prb,twocolum,superscriptaddress,bibnotes,amsmath,amssymb,floatfix, dvipsnames]{revtex4-1}
\usepackage[colorlinks=true,citecolor=blue,linkcolor=magenta]{hyperref}

\usepackage[markup=nocolor, authormarkupposition=left]{changes} 
\usepackage{soul}
\usepackage[utf8]{inputenc}
\usepackage[english]{babel}
\usepackage{amsmath,amsfonts,amssymb}
\usepackage[T1]{fontenc}
\usepackage{url}

\usepackage{amsmath}
\usepackage{amsfonts}
\usepackage{amssymb}

\usepackage{epstopdf}
\usepackage{graphicx}
\graphicspath{{./Figures/}}

\usepackage{changes}

\begin{document}
\title{A wideband, high-resolution vector spectrum analyzer for integrated photonics}

\author{Yi-Han Luo}
\thanks{These authors contributed equally to this work.}
\affiliation{International Quantum Academy, Shenzhen 518048, China}
\affiliation{Shenzhen Institute for Quantum Science and Engineering, Southern University of Science and Technology,
Shenzhen 518055, China}

\author{Baoqi Shi}
\thanks{These authors contributed equally to this work.}
\affiliation{International Quantum Academy, Shenzhen 518048, China}
\affiliation{Department of Optics and Optical Engineering, University of Science and Technology of China, Hefei, Anhui 230026, China}

\author{Wei Sun}
\affiliation{International Quantum Academy, Shenzhen 518048, China}

\author{Ruiyang Chen}
\affiliation{International Quantum Academy, Shenzhen 518048, China}
\affiliation{Shenzhen Institute for Quantum Science and Engineering, Southern University of Science and Technology,
Shenzhen 518055, China}

\author{Sanli Huang}
\affiliation{International Quantum Academy, Shenzhen 518048, China}
\affiliation{Hefei National Laboratory, University of Science and Technology of China, Hefei 230088, China}

\author{Zhongkai Wang}
\affiliation{International Quantum Academy, Shenzhen 518048, China}

\author{Jinbao Long}
\affiliation{International Quantum Academy, Shenzhen 518048, China}

\author{Chen Shen}
\affiliation{International Quantum Academy, Shenzhen 518048, China}

\author{Zhichao Ye}
\affiliation{Qaleido Photonics, Hangzhou 310000, China}

\author{Hairun Guo}
\affiliation{Key Laboratory of Specialty Fiber Optics and Optical Access Networks, Shanghai University, Shanghai 200444, China}

\author{Junqiu Liu}
\email[]{liujq@iqasz.cn}
\affiliation{International Quantum Academy, Shenzhen 518048, China}
\affiliation{Hefei National Laboratory, University of Science and Technology of China, Hefei 230088, China}

\hyphenation{OVNAs}

\maketitle

\noindent\textbf{The analysis of optical spectra -- emission or absorption -- has been arguably the most powerful approach for discovering and understanding matters. 
The invention and development of many kinds of spectrometers have equipped us with versatile yet ultra-sensitive diagnostic tools for trace gas detection, isotope analysis, and resolving hyperfine structures of atoms and molecules.
With proliferating data and information, urgent and demanding requirements have been placed today on spectrum analysis with ever-increasing spectral bandwidth and frequency resolution. 
These requirements are especially stringent for broadband laser sources that carry massive information, and for dispersive devices used in information processing systems.
In addition, spectrum analyzers are expected to probe the device's phase response where extra information is encoded. 
Here we demonstrate a novel vector spectrum analyzer (VSA) that is capable of characterizing passive devices and active laser sources in one setup. 
Such a dual-mode VSA can measure loss, phase response and dispersion properties of passive devices. 
It also can coherently map a broadband laser spectrum into the RF domain. 
The VSA features a bandwidth of 55.1 THz (1260 to 1640 nm), frequency resolution of 471 kHz, and dynamic range of 56 dB. 
Meanwhile, our fiber-based VSA is compact and robust. 
It requires neither high-speed modulators and photodetectors, nor any active feedback control.
Finally, we successfully employ our VSA for applications including characterization of integrated dispersive waveguides, mapping frequency comb spectra, and coherent light detection and ranging (LiDAR). 
Our VSA presents an innovative approach for device analysis and laser spectroscopy, and can play a critical role in future photonic systems and applications for sensing, communication, imaging, and quantum information processing. 
}


\begin{figure*}[t!]
\centering
\includegraphics{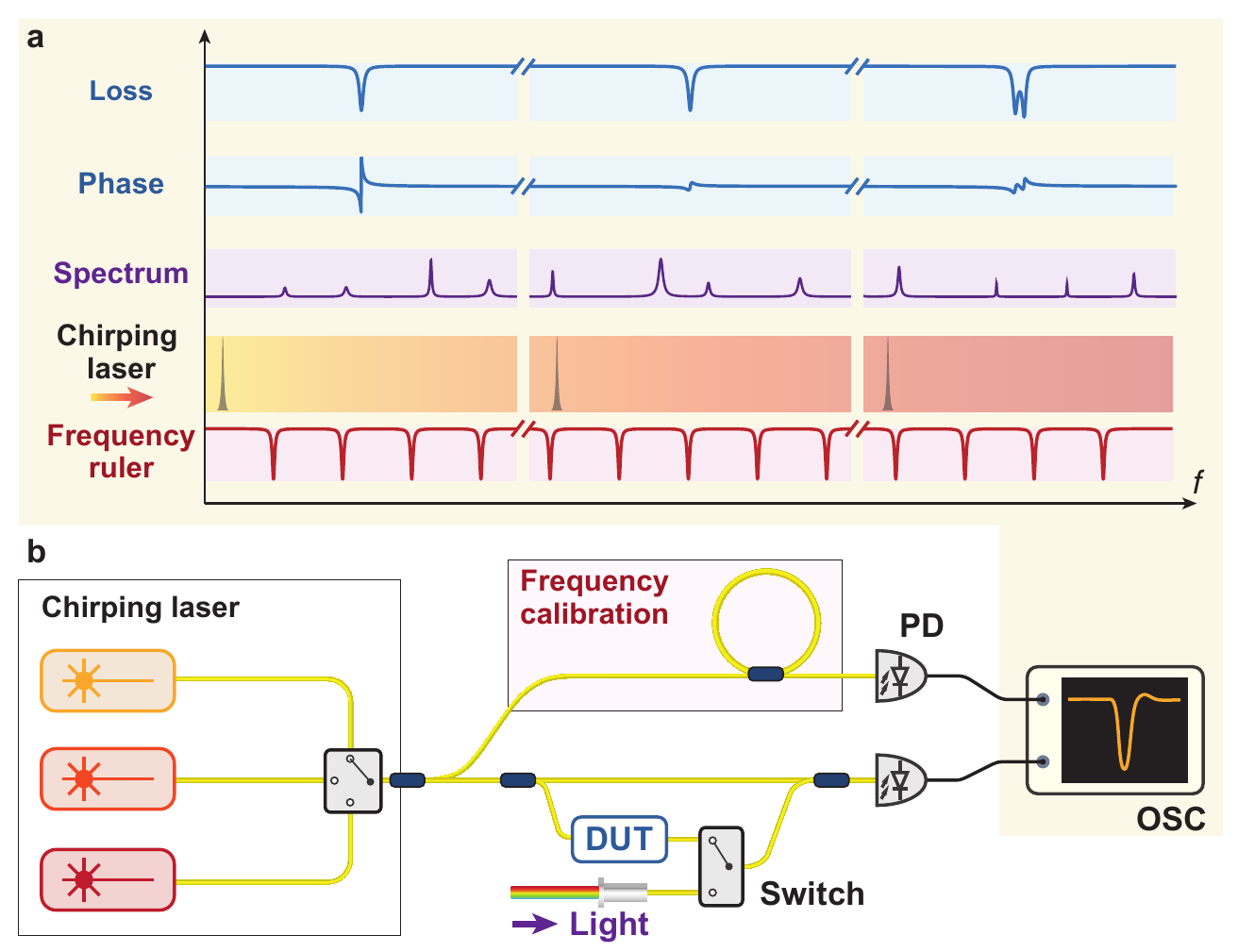}
\caption{
\textbf{Principle and architecture of the vector spectrum analyzer (VSA)}. 
\textbf{a}. 
The principle of our VSA is based on a chirping CW laser that is sent to and transmits through a device under test (DUT). 
The DUT can be either a passive device or a broadband laser source.
The transmission spectrum of the chirping laser through the DUT is a time-domain trace. 
For passive devices, this trace carries the information of the DUT's loss, phase and dispersion over the chirp bandwidth.
For active laser sources, the chirping laser beats progressively with different frequency components of the optical spectrum, thus analyzing the beat signal in the RF domain allows extraction of the spectral information. 
In short, the chirping laser coherently maps the DUT's frequency-domain response into the time domain.
Critical to this frequency-time mapping is precise and accurate calibration of the instantaneous laser frequency during chirping.
This requires to refer the chirping laser to a ``frequency ruler''.
\textbf{b}. 
Experimental setup. 
The frequency-calibration unit here is a phase-stable fiber cavity of 55.58 MHz FSR. 
The chirping laser unit can be a single laser, or multiple lasers that are bandwidth-cascaded together.
The latter allows the extension of the full spectral bandwidth by seamless stitching of individual laser traces into one trace. 
PD, photodetector. 
OSC, oscilloscope. 
}
\label{Fig:1}
\end{figure*}

\noindent \textbf{Introduction}. 
The analysis of light and its propagation in media is fundamental in our information society. 
The discovery of light refraction and dispersion in media has resulted in the invention of prisms and gratings that are ubiquitously used in today's optical systems for imaging, sensing, and communication.
Key enabling building blocks to these applications are dispersive elements that separate light components of different colors (i.e. frequencies) either spatially or temporally \cite{YangZY:21}, with precisely calibrated chromatic dispersion. 
With these elements, modern optical spectrum analyzers (OSA) and spectrometers can deliver unrivaled frequency resolution, large dynamic range, and wide spectral bandwidth of hundreds of nanometers.
Time-stretched systems \cite{Mahjoubfar:17} can probe ultrafast and rare events in complex nonlinear systems.

For spectrum analysis, precise and broadband frequency-calibration of dispersive elements is pivotal. 
Due to the ultimate need for spectrometers with reduced size, weight, cost, and power consumption, extensive effort have been made to create miniaturized spectrometers \cite{Li:22, Ryckeboer:13, Pohl:20, YangZY:19, Yoon:22, Zhang:22} and broadband laser sources \cite{Kippenberg:18, Gaeta:19, Diddams:20, Moss:13, Shu:22, Lu:20, Perez:23, Li:16, Lu:19a, Johnson:15, Porcel:17a, Carlson:17a, Guo:18, Pu:18,Ye:21a, Riemensberger:22} based on integrated waveguides.
For these devices, frequency-calibration is particularly crucial yet challenging since the dispersion of integrated waveguides can be significantly altered by the structures and sizes \cite{Foster:04}. 
Meanwhile, stationary phase approximation for time-stretch dispersive Fourier transform \cite{Goda:13} necessitates carefully frequency-calibrated elements that are strongly dispersive.  
For these purposes, optical vector network analyzers (OVNA) are viable tools. 
Analog to an electrical VNA, an OVNA enables direct characterization of the linear transfer function (LTF) of passive devices, therefore allowing simultaneous measurement of transmission (i.e. loss),  phase response, and dispersion \cite{Sandel:98, Gifford:05, Jin:13}.  
Previously demonstrated OVNAs are based on interferometry \cite{Gifford:05, Li:12}, optical channel estimation \cite{Yi:12, Jin:13}, single-sideband modulation \cite{Tang:12, Pan:17}, and frequency-comb-assisted asymmetric double sidebands \cite{Qing:19}.
Despite this, all these methods have limited measurement bandwidth of sub-terahertz to a few terahertz. 
Therefore for booming demands to understand and to engineer devices used for broadband laser sources that span over tens of terahertz, including optical frequency combs \cite{Kippenberg:18, Gaeta:19, Diddams:20}, parametric oscillators \cite{Moss:13, Lu:20, Perez:23}, quantum frequency translators \cite{Li:16, Lu:19a}, supercontinua \cite{Johnson:15, Porcel:17a, Carlson:17a, Guo:18}, and parametric amplifiers \cite{Pu:18,Ye:21a, Riemensberger:22},  all these methods fail.  

Here we demonstrate a new paradigm of vector spectrum analysis that units OVNA for passive devices and OSA for active laser sources in one setup. 
Our vector spectrum analyzer (VSA) can measure LTF and dispersion property of passive devices, or coherently map an optical spectrum into the RF domain. 

\section{Principle and setup}

\begin{figure*}[t!]
\centering
\includegraphics{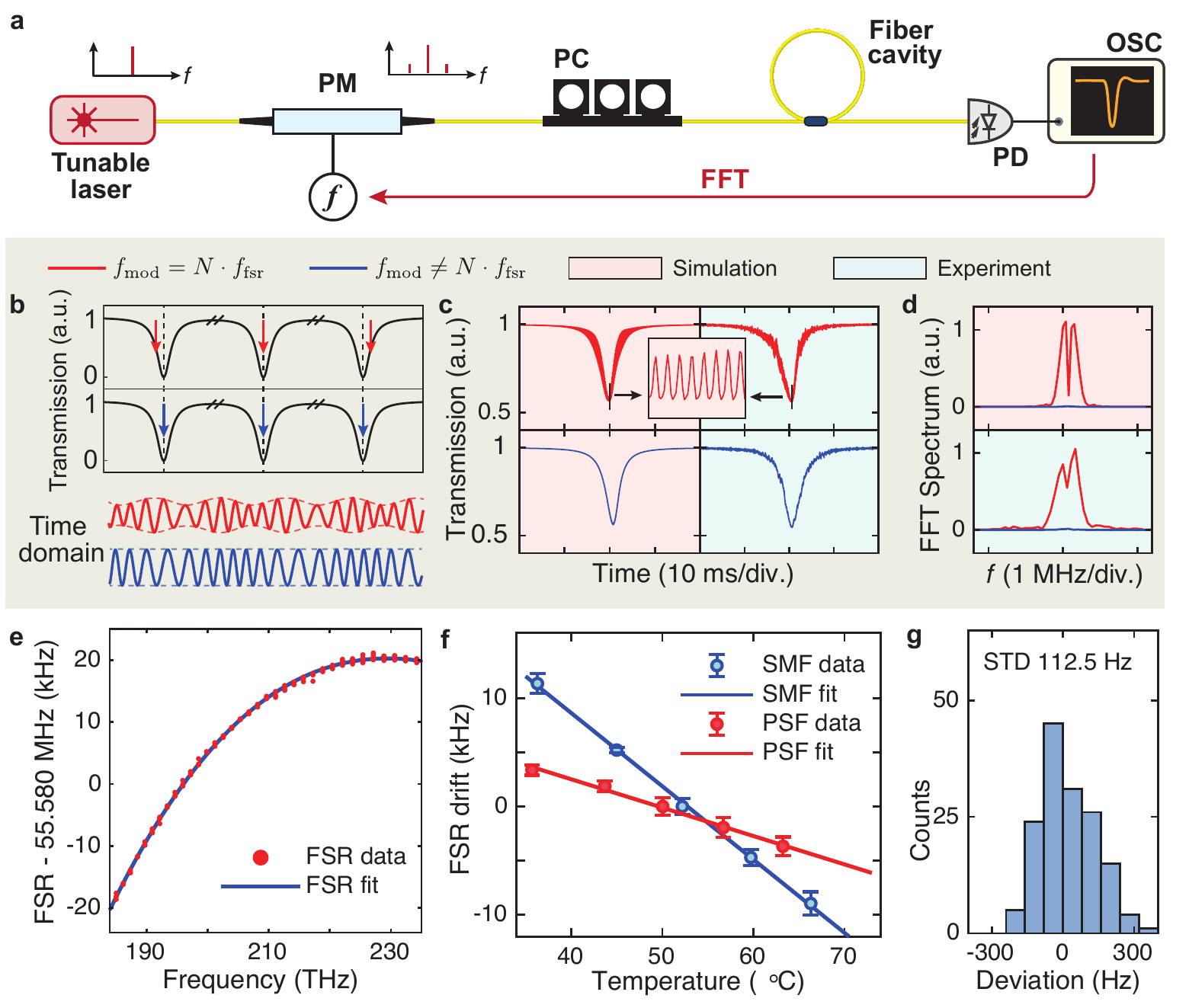}
\caption{
\textbf{Frequency-calibration of the fiber cavity}. 
\textbf{a}. Experimental setup. 
PM, phase modulator. 
PC, polarization controller.
PD, photodetector. 
OSC, oscilloscope. 
FFT, fast Fourier Transformation.
\textbf{b-d}. 
Principle of frequency-calibration process of the fiber cavity's FSR. 
Charts compare the differences when $f_\text{mod}\neq N\cdot f_\text{fsr}$ (red curves) and $f_\text{mod}=N\cdot f_\text{fsr}$ (blue curves). 
From the experimental data (blue area) and simulation (red area), the differences are revealed in the envelope modulation on the time-domain trace (panel b), fiber cavity's resonance profile (panel c), and Fourier peaks in the RF domain (panel d). 
\textbf{e}. 
Measured fiber cavity's FSR variation over the 55.1 THz frequency range with fitted dispersion. 
We perform the measurement at two different temperatures $T_0$ and $T_0+\Delta T$, where $T_0=23^\circ$C and $\Delta T=9.3^\circ$C.
\textbf{f}. 
For fiber cavities made of single-mode fibers (SMF) or phase-stable fibers (PSF), the measured cavity FSR drifts versus relative temperature change, as well as the linear fit. 
\textbf{g}. 
Totally 150 measurements of the fiber cavity's FSR show a standard deviation (STD) of $112.5$ Hz.  
}
\label{Fig:2}
\end{figure*}

The principle of our VSA is illustrated in Fig. \ref{Fig:1}a.
A continuous-wave (CW), widely chirping laser is sent to and transmits through a device under test (DUT) that can be either a passive device or a laser source.
During laser chirping, for passive devices, the frequency-dependent LTF containing the DUT's loss and phase information is photodetected and recorded. 
For laser sources, the chirping laser beats progressively with different frequency components of the optical spectrum, and the beatnote signal is digitally recorded in the RF domain using a narrow-band-pass filter. 
In both cases, the VSA outputs a time-domain trace, with each data point corresponding to the DUT's instantaneous response at a particular frequency during laser chirping. 
In short, the chirping laser coherently maps the DUT's frequency-domain response into the time domain.
Since the laser cannot chirp perfectly linearly, critical to this frequency-time mapping is precise and accurate calibration of the instantaneous laser frequency at any given time.
This requires to refer the chirping laser to a calibrated ``frequency ruler''. 

Following this principle, we construct the setup as shown in Fig. \ref{Fig:1}b. 
A widely tunable, mode-hop-free, external-cavity diode laser (ECDL, Santec TSL) is used as the chirping laser. 
Cascading multiple ECDLs covering different spectral ranges allows the extension of full spectral bandwidth, which is 1260 to 1640 nm (55.1 THz) in our VSA with three ECDLs (see Note 1 in Supplementary Materials). 

The ECDL's CW output is split into two branches. 
One branch is sent to the DUT and the other is sent to a frequency-calibration unit. 
Such frequency-calibration involves relative- (i.e. the frequency change relative to the starting laser frequency) and absolute-frequency-calibration (i.e. accurately measured starting laser frequency), 
The absolute-frequency calibration is performed by referring to a built-in wavelength meter with an accuracy of $200$ MHz (see Note 1 in Supplementary Materials). 
The relative-frequency calibration is described in the following.

\vspace{0.3cm}

\noindent \textbf{Relative-frequency calibration}. 
Figure \ref{Fig:2} illustrates the principle of relative-frequency calibration. 
We use a fiber cavity with an equidistant grid of resonances as the frequency ruler. 
By counting the number of resonances passed by the chirping laser and multiplying the number with the fiber cavity's free spectral range (FSR, $f_\text{fsr}$), the laser frequency excursion (i.e. the instantaneous laser frequency) is calculated.
Extrapolation of laser frequency between two neighbouring fiber cavity's resonances further improves frequency resolution, precision and accuracy, which will be discussed later.
Therefore, critical to this method is the measurement precision of $f_\text{fsr}$ and compensation of fiber dispersion to account $f_\text{fsr}$ variation over the 55.1 THz spectral range.

The experimental setup to calibrate $f_\text{fsr}$ is shown in Fig. \ref{Fig:2}a. 
The ECDL's CW output is phase-modulated by an RF signal generator to create a pair of sidebands. 
The carrier and both sidebands are together sent into the fiber cavity with maintained polarization. 
The transmitted signal through the fiber cavity is probed by a 125-MHz-bandwidth photodetector, analyzed by an oscilloscope, and fed back to the RF signal generator.
Based on the fiber cavity length, an initial value of the fiber cavity's FSR, $\Delta f_0=55.58$ MHz, is estimated. 
The RF driving frequency $f_\text{mod}$ of the phase modulator is set to $f_\text{mod}=N\cdot \Delta f_0$, where $N$ is an integer ($N=3$ in our case). 
Since $\Delta f_0 \neq f_\text{fsr}$, as shown in Fig. \ref{Fig:2}b, in the frequency domain, the carrier and both sidebands locate at different positions of the respective three resonances. 
Therefore, the three CW components experience different cavity responses, and together create an amplitude interference in the time domain at the fiber cavity's output.
This interference can be completely eliminated when $f_\text{mod}$ is slightly varied such that $f_\text{mod}=N\cdot f_\text{fsr}$ is satisfied. 

This time-domain interference can be photodetected and observed by the oscilloscope. 
Figure \ref{Fig:2}c depicts the transmission spectrum of a cavity resonance. 
When $f_\text{mod}\neq N\cdot f_\text{fsr}$, the resonance profile is modulated (red curves); 
When $f_\text{mod}=N\cdot f_\text{fsr}$, the resonance profile is unaffected as a normal Lorentzian profile probed by a single CW laser (blue curves). 
We also simulate this modulation behavior (left red panels) which agrees with the experimental data (right blue panels). 
The modulation amplitude is extracted with fast Fourier transformation (FFT) as shown in Fig. \ref{Fig:2}d, where red curves represent $f_\text{mod}\neq N\cdot f_\text{fsr}$ and blue curves represent $f_\text{mod}=N\cdot f_\text{fsr}$.
When $f_\text{mod}\neq N\cdot f_\text{fsr}$, a binary search to minimize $|f_\text{mod}-N\cdot \Delta f_0|$ is performed, until the modulation peaks vanishes, signalling $f_\text{mod}=N\cdot f_\text{fsr}$. 

We apply this method to measure the fiber cavity's $f_\text{fsr}$ from 1260 to 1640 nm wavelength (55.1 THz frequency range) with an interval of 10 nm, with ambient temperature of $T_0=23.5~^\circ$C. 
The fiber cavity is made of phase-stable fibers (PSF, described later). 
Figure \ref{Fig:2}e shows that, plots and analysis of frequency-dependent $f_\text{fsr}$ enable extraction of the fiber dispersion using a cubic polynomial fit (see Note 2 in Supplementary Materials).
This dispersion-calibrated fiber cavity's resonance grid is used as the frequency ruler in our VSA and following experiments. 

We further characterize the temperature stability of $f_\text{fsr}$. 
The fiber cavity is heated and its $f_\text{fsr}$ shift versus the relative temperature change at 1490 nm is measured, as shown in Fig. \ref{Fig:2}f. 
In addition, we compare two types of fibers to construct the cavity: the normal single-mode fiber (SMF, blue data) and phase-stable fiber (PSF, red data). 
The linear fit shows that the PSF-based fiber cavity features temperature-sensitivity of $\text{d} f_\text{fsr}/\text{d}T=-262$~Hz/K, in comparison to $-676$~Hz/K of the SMF. 
The lower $\text{d} f_\text{fsr}/\text{d}T$ of PSF is the reason why we use PSF instead of SMF. 
Correspondingly, 1 K temperature change (the level of our ambient temperature stabilization and control) causes $\sim 240$ MHz cumulative error of the PSF-based fiber cavity over the entire $55.1$ THz bandwidth. 

We also measure the fiber cavity's dispersion at elevated temperature $T_0+\Delta T$, where $\Delta T=9.3~^\circ$C. 
Figure \ref{Fig:2}e shows that, the two measured fiber dispersion curves at different temperatures are nearly identical except with a global relative shift in the y-axis. 
This indicates that the temperature change only affects $f_\text{fsr}$ but not fiber dispersion. 
More details concerning the measurement are found in Note 3 in Supplementary Materials. 
Therefore, once the ambient temperature is known, the $f_\text{fsr}$ at 1490 nm can be calculated, as well as the $f_\text{fsr}$ variation over frequency. 

Finally, to verify the measurement reproducibility, the $f_\text{fsr}$ value at 1490 nm is repeatedly measured 150 times. 
Figure~\ref{Fig:2}g shows the occurrence histogram, with a standard deviation of $112.5$ Hz. 

Here we use a dispersion-calibrated, phase-stable fiber cavity for relative-frequency calibration. 
We note that frequency comb spectrometers \cite{Diddams:20, Coddington:16, Yang:19} with a precisely equidistant grid of frequency lines can also be used \cite{DelHaye:09, Twayana:21}. 
While frequency combs have been a proven technology for spectroscopy \cite{Picque:19} with unparalleled accuracy, they have several limitations in the characterization of passive devices. 
First, in addition to being bulky and expensive, commercial fiber laser combs as spectrometers suffer from limited frequency resolution due to the RF-rate comb line spacing (typically above 100 MHz). 
Second, the simultaneous injection of more than $10^5$ comb lines can saturate or blind the photodetector, yielding a severely deteriorated signal-to-noise ratio (SNR) and dynamic range.  

Different from frequency combs, CW lasers featuring high photon flux and ever-increasing frequency tunability and agility are particularly advantageous for sensing \cite{Giorgetta:10}. 
In our method, after frequency-calibration by the fiber cavity, the chirping CW laser behaves as a frequency comb with a ``moving'' narrow-band-pass filter, where the filter selects only one comb line each time and rejects other lines. 
Therefore the nearly constant laser power during chirping provides a flat power envelope over the entire spectral bandwidth. 
Therefore our method avoids photodetector saturation and device damage, and increases SNR and dynamic range. 

To improve frequency resolution, the extrapolation of instantaneous laser frequency between two neighbouring fiber cavity's resonances is performed, which relies on the frequency linearity of the chirping laser. 
Such linearity is experimentally characterized in a parallel work \cite{Shi:23} of ours, where the chirping ECDL (Santec TSL) is referenced to a commercial optical frequency comb. 
The result from Ref. \cite{Shi:23} evidences that, using a fiber cavity with 55 MHz FSR and laser chirp rate of 50 nm/s, we experimentally achieve relative-frequency calibration with precision better than 200 kHz. 
The error is caused by the laser chirp nonlinearity.
More details are elaborated in the Note 4 in Supplementary Materials. 
 
The ultimate frequency resolution of each individual time-domain trace is determined by the chirp range divided by the oscilloscope's memory depth ($2\times10^8$). 
For the ECDL of the widest spectral range from 1480 to 1640 nm (19.8 THz), we estimate that the ultimate frequency sample resolution of our VSA is around 99 kHz, i.e. the frequency interval between two recorded neighbouring data points.
The actual resolution can be compromised further by the chirping laser linewidth. 
Therefore we experimentally measured the dynamic laser linewidth using a self-delayed heterodyne setup. 
Experimental details are elaborated in Note 5 in Supplementary Materials.
Within 100 $\mu$s time scale, the ECDL's dynamic linewidth at 50 nm/s chirp rate is averaged as 471 kHz. 
This linewidth is due to multiple reasons including laser intrinsic linewidth, laser chirp nonlinearity, and the fiber delay-line's instability in the heterodyne setup. 
The measured laser dynamic linewidth of 471 kHz sets the lower bound of our VSA's frequency resolution. 

\section{Characterization of passive integrated devices}

\begin{figure*}[t!]
\centering
\includegraphics[width=0.8\textwidth]{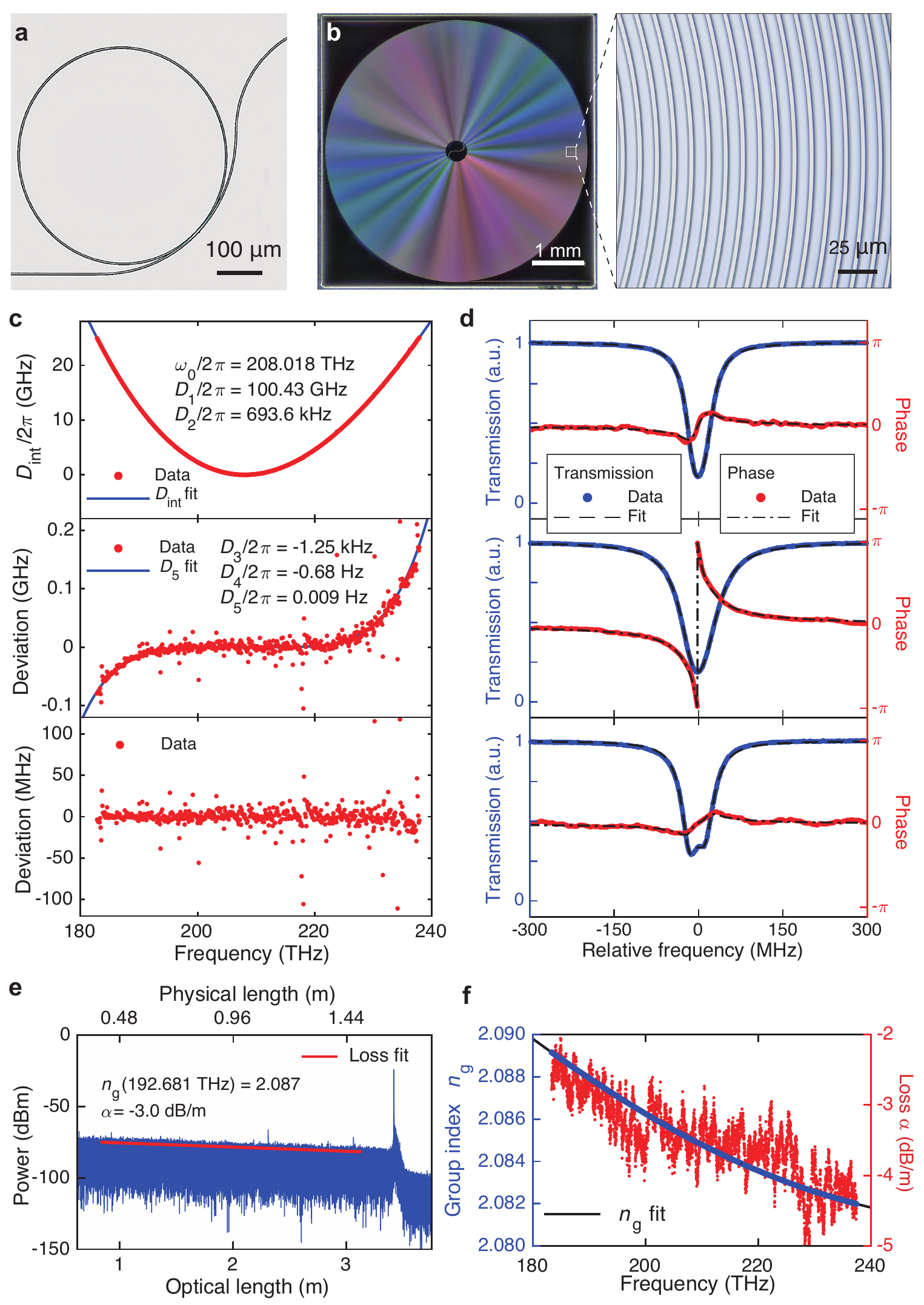}
\caption{
\textbf{Characterization of passive Si$_3$N$_4$ integrated devices}. 
\textbf{a, b}. 
Optical microscope images showing a microresonator coupled with a bus waveguide (panel a), and a 1.6394-meter-long spiral waveguide contained in a photonic chip of $5\times5$~mm$^2$ size (panel b). 
The zoom-in shows the densely coiled waveguide. 
\textbf{c}. 
Measured integrated microresonator dispersion profile and fit up to the fifth order. 
\textbf{d}. 
Measured transmission and phase profiles of three resonances that are under-coupled (top), over-coupled (middle), or feature mode split (bottom, also under-coupled).
\textbf{e}. 
Measured OFDR data of the spiral waveguide. 
The major peak at 1.6394 meter physical length (3.4214 meter optical length) is attributed to the light reflection at the rear chip facet, where the waveguide terminates. 
This length difference indicates a group index $n_g=2.087$ at 192.681 THz.
The loss rate $\alpha=-3.0$ dB/m (physical length) is calculated with a linear fit of power decrease over distance (red line).
\textbf{f}. 
Measured group index $n_g$ (blue dots) and loss $\alpha$ (red dots) of the waveguide over the 55.1 THz spectral range.
}
\label{Fig:3}
\end{figure*}

Next we demonstrate several applications using our VSA. 
We first use our VSA as an OVNA to characterize passive devices. 
We select two types of optical devices: an integrated optical microresonator and a meter-long spiral waveguide. 
Both devices, fabricated on silicon nitride (Si$_3$N$_4$) \cite{Ye:23}, have been extensively used in integrated nonlinear photonics \cite{Moss:13, Gaeta:19}. 
For example, optical microresonators of high quality ($Q$) factors are central building blocks for miniaturized microresonator-soliton-based optical frequency combs (``soliton microcomb'') \cite{Kippenberg:18, Gaeta:19, Diddams:20}, ultralow-threshold optical parametric oscillators \cite{Moss:13, Lu:20, Perez:23}, and quantum frequency translators \cite{Li:16, Lu:19a}. 
Ultralow-loss, dispersion-flattened waveguides are cornerstones for multi-octave supercontinua \cite{Johnson:15, Porcel:17a, Carlson:17a, Guo:18} and continuous-travelling-wave optical parametric amplifiers \cite{Pu:18,Ye:21a, Riemensberger:22}. 
All these applications require precisely characterized properties of integrated devices, such as loss, phase, and dispersion over a bandwidth spanning more than 100 nm. 

\vspace{0.3cm}

\noindent \textbf{Characterization of integrated optical microresonators}.
Figure \ref{Fig:3}a shows an optical microscope image of a Si$_3$N$_4$ optical microresonator. 
The resonance frequency $\omega/2\pi$ and linewidth $\kappa/2\pi$ of each fundamental-mode resonances, ranging from 1260 nm (237.9 THz) to 1640 nm (182.8 THz) wavelength, are measured. 
The microresonator's integrated dispersion is defined as 
\begin{equation}
D_\text{int}(\mu)=\omega_\mu-\omega_0-D_1\mu=\sum_{n=2}^{\cdots}\frac{D_n\mu^n}{n!}
\label{Eq.Dint}
\end{equation}
where $\omega_{\mu}/2\pi$ is the $\mu$-th resonance frequency relative to the reference resonance frequency $\omega_0/2\pi$, 
$D_1/2\pi$ is microresonator FSR, 
$D_2/2\pi$ describes group velocity dispersion (GVD), 
and $D_3$, $D_4$, $D_5$ are higher-order dispersion terms. 
Figure \ref{Fig:3}c top plots the measured $D_\text{int}$ profile, with each parameter extracted from the fit using Eq. \ref{Eq.Dint}. 
We note that, due to our 55.1 THz measurement bandwidth and 471 kHz frequency resolution, our method can measure higher-order dispersion  \cite{YangK:16} up to the fifth-order $D_5$ term. 
This is validated in Fig. \ref{Fig:3}c middle, where $D_2$, $D_3$ and $D_4$ terms are subtracted from $D_\text{int}$, and the residual dispersion is fitted with $D_5\mu^5/120$. 
Figure \ref{Fig:3}c bottom shows that, after further subtraction of the $D_5$ term, no prominent residual dispersion is observed. 
Some data points deviate from the fit due to avoided mode crossings in the microresonator \cite{Herr:14a}.

For each resonance fit \cite{Li:13}, the intrinsic loss $\kappa_0/2\pi$, external coupling strength $\kappa_\text{ex}/2\pi$, and the total (loaded) linewidth $\kappa/2\pi=(\kappa_0+\kappa_\text{ex})/2\pi$, are extracted. 
Figure \ref{Fig:3}d shows three typical resonances with fit curves (blue), including one with visible mode split (bottom). 
Conventionally, based on a single resonance profile, it is impossible to judge whether the resonance is over-coupled ($\kappa_\text{ex}>\kappa_0$) or under-coupled ($\kappa_\text{ex}<\kappa_0$) \cite{Cai:00}. 
The coupling condition can only be revealed by phase (vector) measurement. 

Here we split laser power into two branches as shown in Fig. \ref{Fig:1}b. 
In one branch the laser transmits through the DUT, while in the other the laser experiences a delay $\Delta \tau$. 
The delay $\Delta \tau$ introduces a frequency difference $\Delta f=\gamma\Delta\tau$ between the two branches, where $\gamma$ is the laser chirp rate. 
Thus when the two branches recombine, a beat signal is photodetected. 
The extra phase shift $\phi$ introduced by the DUT also applies to the beat signal, which can be extracted with Hilbert transformation \cite{Marple:99} (see Note 6 in Supplementary Materials). 
The measured and fitted phases are shown in Fig. \ref{Fig:3}d red curves. 
The continuous phase transition across the resonance in Fig. \ref{Fig:3}d top and bottom represents under-coupling, while the phase jump by 2$\pi$ in Fig. \ref{Fig:3}d middle represents over-coupling.
From top to bottom, the fitted loss values $(\kappa_0/2\pi, \kappa_\text{ex}/2\pi)$ for each resonances are $(23.8, 14.0)$,  $(19.9, 42.4)$, and $(24.7, 12.8)$ MHz. 
The complex coupling coefficient \cite{Pfeiffer:18} in the bottom is $\kappa_c/2\pi=29.1+2.25i$ MHz.

\vspace{0.3cm}

\noindent \textbf{Characterization of single-pass waveguides}.
In addition to microresonators as well as other resonant structures, our method can also characterize single-pass waveguides. 
Figure \ref{Fig:3}b shows an optical microscope image of a Si$_3$N$_4$ photonic chip containing a spiral waveguide of $L_0=1.6394$ meter physical length.
We use our VSA as optical frequency-domain reflectometry (OFDR) \cite{Soller:05} to characterize the waveguide loss and dispersion.
Figure \ref{Fig:3}e plots the OFDR signal from the spiral waveguide. 
The prominent peak located at 1.6394 meter physical length (3.4214 meter optical length) is attributed to the light reflection at the rear facet of the chip, where the waveguide terminates. 
The difference in the physical and optical lengths indicates a group index of $n_g=2.087$ at 192.681~THz. 

In the presence of waveguide dispersion, the optical path length $L_\text{op}$ varies due to the frequency-dependent $n_g$.
This dispersion-induced optical path variation leads to deteriorated spatial resolution in broadband measurement \cite{Glombitza:93}. 
By dividing the broadband measurement data into narrow-band segments \cite{Bauters:11, Zhao:17}, the optical path length at different optical frequencies can be obtained, and thus the frequency-dependent $n_g$ over the $55.1$ THz spectral range can be extracted.
With the extracted $n_g$, the waveguide dispersion can be de-embedded with a re-sample algorithm \cite{Kohlhaas:91, Zhao:17}.

Light traveling in the waveguide experiences attenuation following the Lambert-Beer Law $I(L)=I_0\cdot\text{exp}(\alpha L)$. 
In Fig. \ref{Fig:3}e, the average linear loss $\alpha=-3.0$ dB/m (physical length) is extracted by applying a first-order polynomial fit of the power profile (red line) within the $19.8$ THz bandwidth and centered at $192.681$ THz.
Figure \ref{Fig:3}f shows the frequency-dependent $\alpha$ (red dots) and $n_g$ (blue dots) extracted using segmented OFDR algorithm \cite{Bauters:11, ZhaoQ:20}. 
The $n_g$ is further fitted at 208.015 THz, and the dispersion parameters are extracted up to the fourth order as $\beta_1=6955.0$ fs/mm, $\beta_2=-74.09$ fs$^2$/mm, $\beta_3=199$ fs$^3$/mm, and $\beta_4=2.4\times10^{2}$ fs$^4$/mm.
The loss fluctuation with varying frequency is likely due to multi-mode interference in the spiral waveguide \cite{Ji:22}. 

In OFDR, the resolution $\delta L_\text{op}$ of optical path length is determined by the laser chirp bandwidth $B$ as $\delta L_\text{op}=c/2B$, with $c$ being the speed of light in vacuum. 
Our VSA can provide a maximum $B=19.8$ THz in a single measurement, which enables $\delta L_\text{op}=7.6$ $\mu$m.
As shown in Fig. \ref{Fig:3}e, such a fine resolution allows unambiguous discrimination of scattering points in the waveguide, which are revealed by small peaks.
Thus our VSA is proved as a useful diagnostic tool for integrated waveguides.

\begin{figure*}[t!]
\centering
\includegraphics[width=0.8\textwidth]{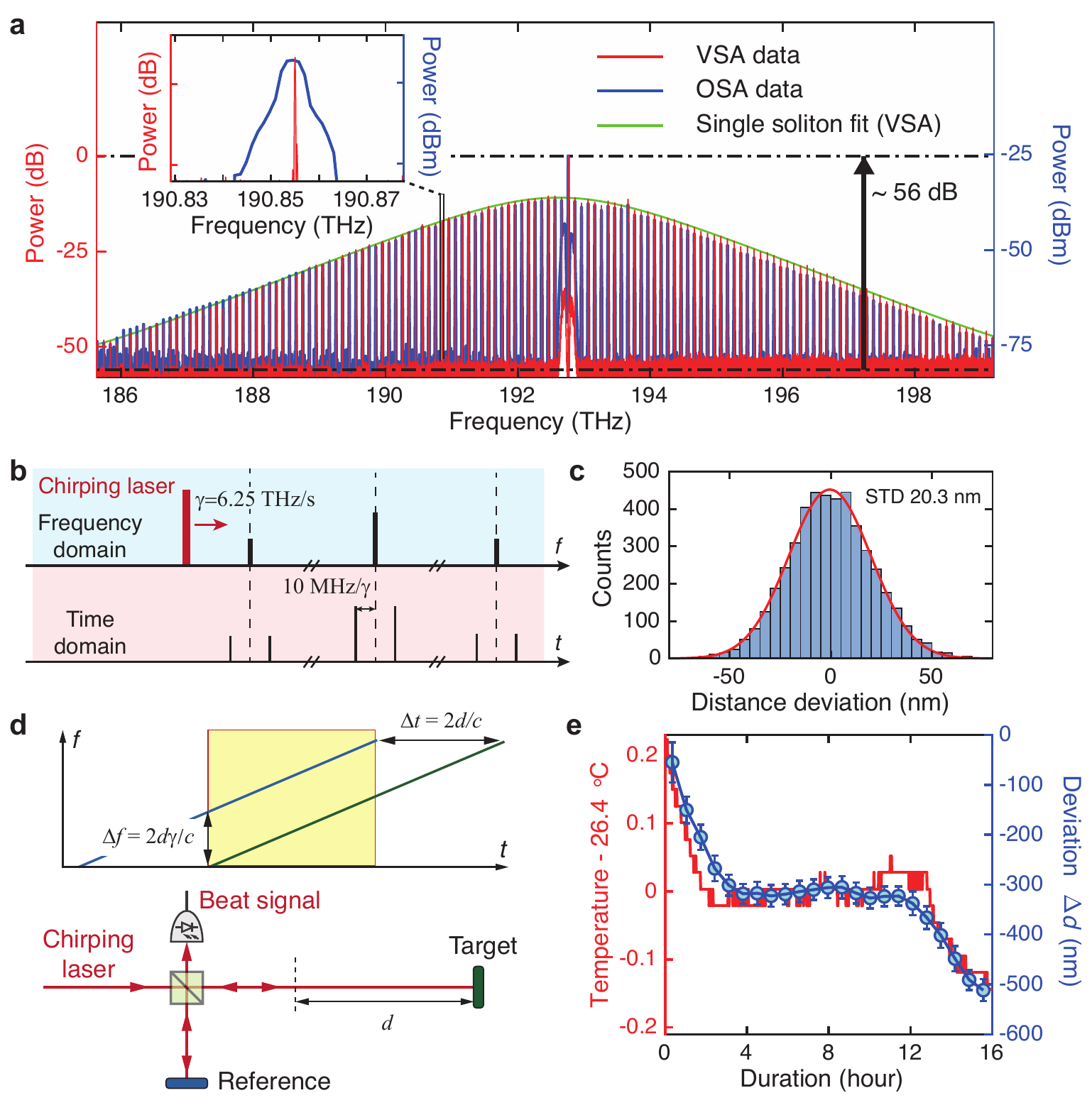}
\caption{
\textbf{Characterization of broadband laser spectra and coherent LiDAR applications}. 
\textbf{a}. 
Single soliton spectra measured by our VSA (red) and a commercial OSA (blue). 
The spectral envelope of VSA data is fitted with a sech$^2$ function (green). 
Inset: Zoom-in of the comb line resolved by our VSA and the OSA, demonstrating the significant resolution enhancement by the VSA. 
\textbf{b}. 
Principle of coherent detection of broadband laser spectra using a chirping laser. 
The laser beats progressively with different frequency components of the optical spectrum, which allows frequency detection in the RF domain and continuous information output in the time domain. 
\textbf{c}. 
Histogram showing the deviations of 4625 LiDAR measurements from their mean values. 
The LiDAR precision is revealed by the standard deviation of 20.3 nm. 
\textbf{d}. 
Principle of coherent LiDAR using a frequency-calibrated chirping laser.
With known chirp rate $\gamma$, the heterodyne measurement of frequency beat in the RF domain $\Delta f=2d\gamma/c$ allows calculation of the time delay $\Delta t=2d/c$ and thus to calculate the distance $d$.
\textbf{f}. 
LiDAR Measurement of thermal expansion of our optical table using our VSA, in comparison with data from a digital ambient thermometer. 
}
\label{Fig:4}
\end{figure*}

\section{Characterization of soliton spectra and LiDAR applications}

\noindent \textbf{Coherent detection of frequency comb spectra}. 
Next, we use our VSA as an OSA to characterize broadband laser spectra.
While modern OSAs can achieve wide spectral bandwidth, they suffer from a limited frequency resolution ranging from sub-gigahertz to several gigahertz. 
This issue prohibits OSAs from resolving fine spectral features.  
For example, individual lines of mode-locked lasers or supercontinua with repetition rates in the RF domain cannot be resolved by OSAs. 
Soliton microcombs with terahertz-rate repetition rate can be useful for low-noise terahertz generation \cite{Tetsumoto:21, Wang:21}, but their precise comb line spacing can neither be measured by normal photodetectors nor OSAs.

Here we demonstrate that our VSA can act as an OSA which features a 55.1 THz spectral range and megahertz frequency resolution. 
As an example, we measure the repetition rate (line spacing) of a 100-GHz-rate soliton microcomb generated. 
The schematic is depicted in Fig. \ref{Fig:4}b, where the laser chirps across the entire soliton spectrum. 
Every time the laser passes through a comb line, it generates a moving beatnote. 
Using a finite impulse response (FIR) band-pass filter of 10 MHz center frequency and 3 MHz bandwidth, the beatnote creates a pair of marker signals when the laser frequency is $\pm 10$ MHz distant from the comb line. 
The polarization of the soliton spectrum is measured by varying the laser polarization until the beat signal with maximum intensity is observed.
This search procedure of polarization can essentially be programmed and automated.  
Since the instantaneous laser frequency is precisely calibrated, the comb line spacing is extracted by calculating the frequency distance from two adjacent pairs of marker signals. 
With the known laser power and measured marker signals' intensity, the absolute power of each comb line can be calculated.

Figure \ref{Fig:4} compares the measured soliton microcomb spectra using our VSA and a commercial OSA. 
Both spectra are nearly identical, particularly in that the left and right y-axes have identical power scales, which validates our VSA measurement. 
The dynamic range of our VSA is found as 56 dB, which is on par with modern commercial OSAs with the finest resolution (e.g. 45 to 60 dB at 0.02 nm resolution for Yokogawa OSAs). 
Figure \ref{Fig:4}a inset evidences that our VSA indeed provides significantly finer frequency resolution than the OSA. The soliton repetition rate measured by the VSA is $(100.307\pm0.002)$ GHz. 

We emphasize that, here the frequency resolution of our VSA as an OSA is limited by the bandwidth of FIR band-pass filters. 
In digital data processing, we find that 3 MHz FIR bandwidth yields the optimal resolution bandwidth of 3 MHz.  
Experimentally, we verify the resolution bandwidth by phase-modulating a low-noise fiber laser (NKT Koheras) to generate a pair of sidebands of 3 MHz difference to the carrier. 
The carrier and the sidebands are unambiguously resolved using our VSA (See Note 5 in Supplementary Materials). 
The 3 MHz resolution bandwidth is also consistent with the uncertainty of measured soliton repetition rate of 100.307 GHz. 

\vspace{0.3cm}

\noindent \textbf{Light detection and ranging}. 
Finally, we note that the broadband, chirping, and interferometric nature of our VSA also enables coherent LiDAR. 
Frequency-modulated continuous-wave (FMCW) LiDAR is a ranging technique based on frequency-modulated interferometry \cite{Amann:01}, as depicted in Fig \ref{Fig:4}d. 
The chirping laser is split into two arms, with one arm to the reference and the other to the target with a path difference of $d$.
When the reflected signals from both arms recombine at the photodetector, the detected beat frequency is determined as $\Delta f=2d\gamma/c$, where $c$ is the speed of light in air and $\gamma$ is the chirp rate. 
Thus the measurement of $\Delta f$ in the RF domain allows distance measurement of $d$. 
The ranging resolution $\delta d$, i.e. the minimum distance that the LiDAR can distinguish two nearby objects, is limited by the chirp bandwidth $B$ as $\delta d=c/2B$. 
One advantage of our VSA as a FMCW LiDAR is that, our laser can provide maximum $B=19.8$ THz that enables $\delta d=7.6$ $\mu$m.

In our LiDAR experiment, we set the linear chirp rate of $\gamma=6.25$ THz/s and duration of $T=0.4$~s. 
The experimental setup and data analysis procedure of LiDAR are found in Note 7 in Supplementary Materials. 
As a demonstration, we monitor the thermal expansion of our optical table due to ambient temperature drift, as shown in Fig. \ref{Fig:4}e. 
The distance difference between the target mirror and the reference mirror on the table is $d=137.63128$ mm.
The measured distance change $\Delta d$ within 500 nm range agrees with the temperature decrease that causes contraction of the optical table.
After subtracting the global trend, Figure \ref{Fig:4}c shows the histogram of the deviations of 4625 measurements from their mean values. 
Our LiDAR precision is revealed by the standard deviation of 20.3 nm. 
Such a precision is provided by the careful relative-frequency calibration and long-term stability of our VSA. 

\section{Conclusion}

In summary, we have demonstrated a dual-mode VSA featuring 55.1 THz spectral bandwidth, 471 kHz frequency resolution, and 56 dB dynamic range. 
The VSA can operate either as an OVNA to characterize the LTF and dispersion property of passive devices, or as an OSA to characterize broadband frequency comb spectra. 
A comparison of our VSA with other state-of-the-art OSAs and OVNAs is shown in Note 5 of Supplementary Material. 
Our VSA can also perform LiDAR with a distance resolution of 7.6 $\mu$m and precision of 20.3 nm. 
Meanwhile, our VSA is fiber-based, and neither requires high-speed modulators and photodetectors, nor any active feedback control.
Therefore the system is compact, robust, and transportable for field-deployable applications.

There are several aspects to further improve the performance and reduce the complexity of our VSA. 
First, the frequency resolution can be improved by increasing oscilloscope's memory depth, or by sacrificing chirp bandwidth, until the laser noise dominates.
Second, the frequency accuracy can be improved by adding a highly stable reference laser in the system.
When the ECDL scans through the reference laser, the two lasers beat and create a marker in the time-domain trace.
The marker marks the point where the chirping ECDL has an instantaneous frequency as the reference laser's frequency.
Third, more ECDLs can be added into the system, allowing further extension of the spectral bandwidth and operation in other wavelength ranges such as the visible and mid-infrared bands. 
Meanwhile, even ECDLs with mode hopping can be used in our VSA. 
The self-calibration and compensation of mode hopping can be realized by adding a calibrated, large-FSR cavity (e.g. a Si$_3$N$_4$ microresonator of terahertz-rate FSR), in addition to the fine-tooth fiber cavity. 
By measuring the resonance-to-resonance frequency and referring to previously calibrated local FSR of the microresonator, the exact mode hopping range and location can be inferred. 
Adding more calibrated cavities of different FSR values to form a Vernier structure can further enhance the precision and accuracy. 

Besides characterization of passive elements and broadband laser sources for integrated photonics, our VSA can also be applied for time-stretched systems \cite{Mahjoubfar:17}, optimized optical coherent tomography (OCT) \cite{Adler:07}, linearization of FMCW LiDAR \cite{Roos:09}, and resolving fine structures in Doppler-free spectroscopy \cite{Meek:18}.
Therefore our VSA presents an innovative approach for device analysis and laser spectroscopy, and can play a crucial role in future photonic systems and applications for sensing, communication, imaging, and quantum information processing. 

\medskip
\begin{footnotesize}

\noindent \textbf{Funding Information}: 
J. Liu acknowledges support from the National Natural Science Foundation of China (Grant No.12261131503), Shenzhen-Hong Kong Cooperation Zone for Technology and Innovation (HZQB-KCZYB2020050), and from the Guangdong Provincial Key Laboratory (2019B121203002).
Y.-H L. acknowledges support from the China Postdoctoral Science Foundation (Grant No. 2022M721482). 

\noindent \textbf{Acknowledgments}: 
We thank Ting Qing and Jijun He for the fruitful discussion on OVNA, Yuan Chen, Zhiyang Chen and, Huamin Zheng for assistance in the experiment, and Lan Gao for taking the sample photos. 
J. Liu is indebted to Dapeng Yu who provided critical support to this project. 

\noindent \textbf{Author contributions}: 
Y.-H. L., B. S., W. S., Z. W., and J. Long built the experimental setup, with assistance and advice from H. G.. 
Y.-H. L., B. S., W. S., and R. C. performed the frequency-calibration. 
B. S., Y.-H. L., W. S., and S. H. performed the experiments on VSA applications. 
C. S. and Z. Y. fabricated the silicon nitride chips. 
Y.-H. L, B. S. , and J. Liu analysed the data and prepared the manuscript with input from others. 
J. Liu supervised the project.  

\noindent \textbf{Conflict of interest}:
Y. -H. L, B. S. , W. S.  and J. Liu are inventors on a patent application related to this work. 
Others declare no conflicts of interest. 

\noindent \textbf{Data Availability Statement}: 
The code and data used to produce the plots within this work will be released on the repository \texttt{Zenodo} upon publication of this preprint.

\end{footnotesize}
%
\end{document}


\title{Supplementary Materials for: A wideband, high-resolution vector spectrum analyzer for integrated photonics}

\author{Yi-Han Luo}
\affiliation{International Quantum Academy, Shenzhen 518048, China}
\affiliation{Shenzhen Institute for Quantum Science and Engineering, Southern University of Science and Technology,
Shenzhen 518055, China}

\author{Baoqi Shi}
\affiliation{International Quantum Academy, Shenzhen 518048, China}
\affiliation{Department of Optics and Optical Engineering, University of Science and Technology of China, Hefei, Anhui 230026, China}

\author{Wei Sun}
\affiliation{International Quantum Academy, Shenzhen 518048, China}

\author{Ruiyang Chen}
\affiliation{International Quantum Academy, Shenzhen 518048, China}
\affiliation{Shenzhen Institute for Quantum Science and Engineering, Southern University of Science and Technology,
Shenzhen 518055, China}

\author{Sanli Huang}
\affiliation{International Quantum Academy, Shenzhen 518048, China}
\affiliation{Hefei National Laboratory, University of Science and Technology of China, Hefei 230088, China}

\author{Zhongkai Wang}
\affiliation{International Quantum Academy, Shenzhen 518048, China}

\author{Jinbao Long}
\affiliation{International Quantum Academy, Shenzhen 518048, China}

\author{Chen Shen}
\affiliation{International Quantum Academy, Shenzhen 518048, China}

\author{Zhichao Ye}
\affiliation{Qaleido Photonics, Hangzhou 310000, China}

\author{Hairun Guo}
\affiliation{Key Laboratory of Specialty Fiber Optics and Optical Access Networks, Shanghai University, Shanghai 200444, China}

\author{Junqiu Liu}
\affiliation{International Quantum Academy, Shenzhen 518048, China}
\affiliation{Hefei National Laboratory, University of Science and Technology of China, Hefei 230088, China}

\maketitle

\section{Absolute frequency calibration and cascading multiple ECDLs for spectral extension}
\vspace{0.5cm}

\begin{figure}[b!]
\centering
\renewcommand{\figurename}{Supplementary Figure}
\includegraphics[width=0.85\columnwidth]{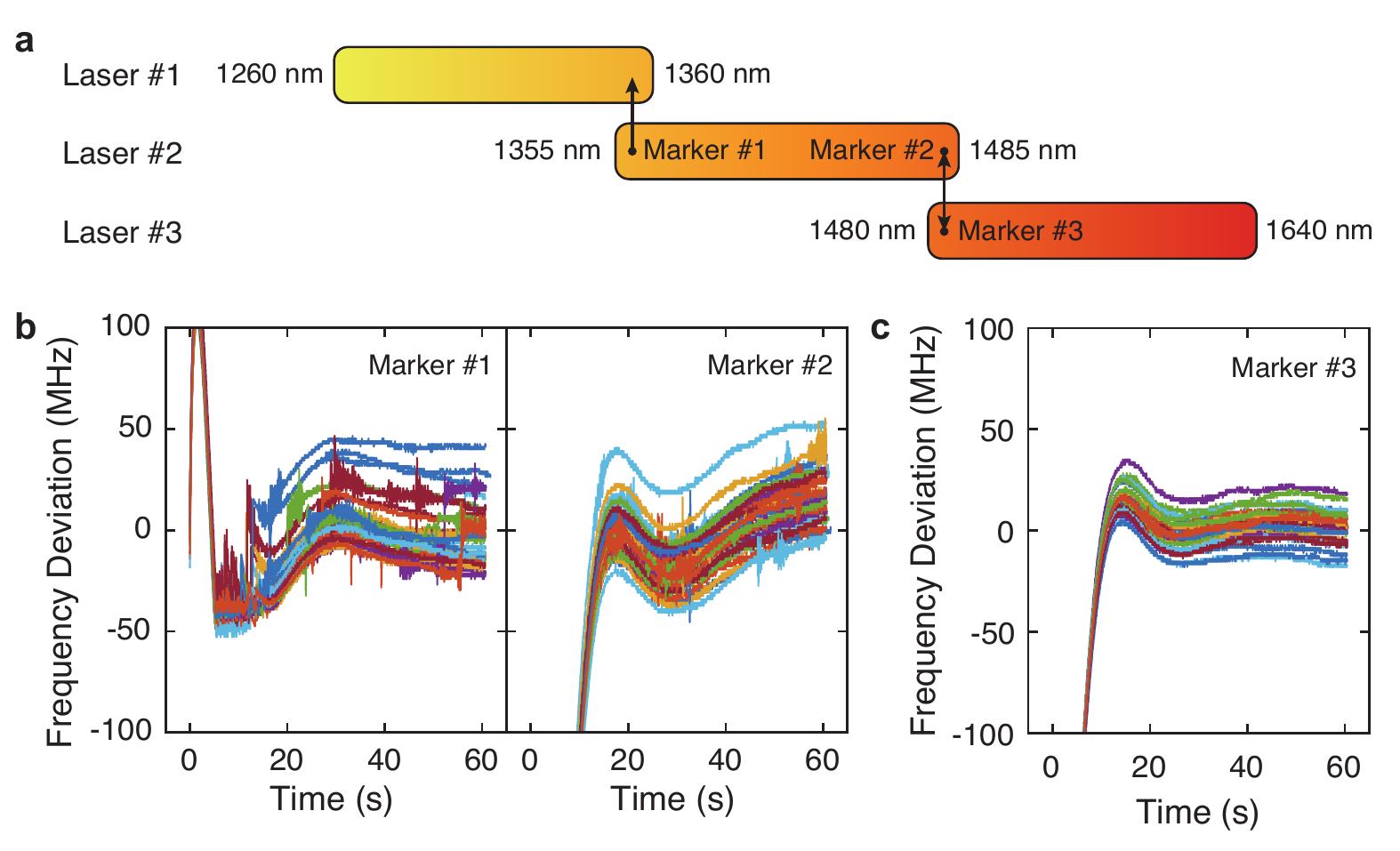}
\caption{
\textbf{laser cascade procedure and absolute frequency calibration.} 
\textbf{a}. 
The spectral ranges of the three ECDLs used in our experiment, and the three selected markers for absolute frequency measurement using a wavelength meter. 
\textbf{b-c}. 
The frequency drift of the markers is measured by the wavelength meter. 
For each data trace, the measurement time duration is 60 s. 
To evidence the marker's frequency deviation, an offset frequency is subtracted. 
The offset frequency is chosen by averaging all the data in all the traces between 30 to 60 seconds. 
}
\label{SIfig:lasercascade}
\end{figure}

In our experiment, three ECDLs are cascaded. 
The respective spectral ranges of the three ECDLs are shown in Fig. \ref{SIfig:lasercascade}a, where each two adjacent ECDLs share a common spectral range.  
In our method, these shared ranges are exploited not only for cascading lasers but also for absolute frequency calibration.  

The Santec ECDL has a monitor panel where the value of the current laser frequency is displayed. 
To set the laser frequency, we send a command to the ECDL via an external computer. 
Then the laser is set to this frequency, and the frequency value is displayed on the laser's monitor panel. 

First, we illustrate the absolute frequency calibration of Laser \#1. 
First, the frequency of Laser \#2 is set to the Marker \#1's frequency at $220.92$ THz (1357.0 nm), as shown in the Fig. \ref{SIfig:lasercascade}a. 
Then Laser \#1 starts to chirp. 
When Laser \#1's frequency scans across Laser \#2's frequency at Marker \#1, the beat signal between Lasers \#1 and \#2 can be detected by the photodetector and recorded by the oscilloscope. 
In our experiment, the beat signal is filtered by a FIR bandpass filter of 10 MHz center frequency. 
In this way, two marker signals are created in the Laser \#1's time trace due to the beat signal, when Laser \#1's frequency is $\pm 10$ MHz detuned to Laser \#2's frequency. 
Therefore the time when Laser \#1 chirps across Marker \#1's frequency (that is known) can be precisely extracted, which acts as an absolute frequency reference for Laser \#1's time trace. 
Similarly, the Marker \#2 set by Laser \#3 is used as the absolute frequency marker of Laser \#2;
The Marker \#3 set by Laser \#2 is used as the absolute frequency marker of Laser \#3.
Using the absolute frequency markers described above, together with the relative frequency calibration using the fiber cavity, the frequency-time mapping of the three chirping ECDLs can be individually constructed. 
Then the three calibrated time traces are seamlessly stitched to form one continuous trace covering the entire 55.1 THz spectral range. 

Next, we experimentally measure the three markers' frequency values with a commercial wavelength meter (HighFinesse WS6-200) with an accuracy of 200 MHz and a resolution of 2 MHz. 
We first test Laser \#2 since Markers \#1 and \#3 are created by Laser \#2. 
The frequency of Laser \#2 is repeatedly switched between 1482.0000 nm ($202.28911$ THz) and 1357.0000 nm ($220.92296$ THz). 
For Marker \#2 created by Laser \#3, the frequency of Laser \#3 is repeatedly switched between 1482.0000 nm ($202.28911$ THz) and 1640.0000 nm ($182.80028$ THz). 
The time interval between two subsequent frequency switches is 60 s.
During this time interval, the laser frequency is tracked and recorded with the wavelength meter.  
The key question to answer is, whether the laser frequencies measured by the wavelength meter are precisely equal to the frequency value we set to the lasers.

The measurement results for Lasers \#2 and \#3 are shown in Fig. \ref{SIfig:lasercascade}(b, c). 
For each marker, a total of 60 traces are measured, and each trace is continuously measured for 60 s. 
In each trace, the data from 30 s to 60 s are used to calculate the mean frequency. 
The mean frequency values of Marker \#1, \#2 and \#3, measured by the wavelength meter, are $220.92340(2)$ THz, $202.28896(2)$ THz and $202.28901(1)$ THz, respectively. 
The differences between the measured frequency values from the set values are $0.44$ GHz, $0.15$ GHz, and $0.10$ GHz. 
Figure \ref{SIfig:lasercascade}(b, c) shows that, for each marker, the deviations of measured frequency values to their mean value are within the accuracy of the wavelength meter (200 MHz). 
Therefore, in the experiment, we use the measured frequency values of $220.92340$ THz, $202.28896$ THz, and $202.28901$ THz as the markers' frequency values. 

The workflow of our experiment procedure is detailed in Table \ref{tab:workflow}.
Before the selected ECDL starts chirping, its reference laser is set to the specified frequency and waits 60 s for stabilization. 
The actual emission frequency of the reference laser is measured by the wavelength meter.
After the 60 s, the laser starts chirping. 
The photodetector probes the laser chirping, and the data is recorded by the oscilloscope.
For the other two ECDLs, the operation procedure is similar. 
Meanwhile, during the 60 s waiting time for laser frequency stabilization,  the recorded data by the oscilloscope can be processed by the computer.
Finally, the three individually measured and calibrated data traces of each ECDL are stitched, forming the full 55.1 THz spectral bandwidth measurement. 

\begin{table}[t!]
\centering
\caption{Experimental workflow of the VSA. }
\label{tab:workflow}
\begin{tabular}{|c|c|c|}
\hline
Time (s) & Laser Action & Data Process Action\\
\hline
0 & Set Laser \#2 at 1357.0000 nm & -\\
60 & Laser \#1 starts chirping & Oscilloscope starts recording data from PD\\
63 & Set Laser\# 3 at 1482.0000 nm & Oscilloscope starts transmitting data to PC\\
88 & - & Frequency calibration of Laser \#1 data trace\\
123 & Laser \#2 starts chirping & Oscilloscope starts recording data from PD\\
126 & Set laser \#2 at 1482.0 nm & Oscilloscope starts transmitting data to PC\\
151 & - & Frequency calibration of Laser \#2 data trace\\
186 & Laser \#3 starts chirping & Oscilloscope starts recording data from PD\\
189 & Finish & Oscilloscope starts transmitting data to PC\\
214 & - & Frequency calibration of Laser \#3 data trace\\
220 & - & Finish\\
\hline
\end{tabular}
\end{table}

\begin{figure}[b!]
\centering
\renewcommand{\figurename}{Supplementary Figure}
\includegraphics[width=0.7\columnwidth]{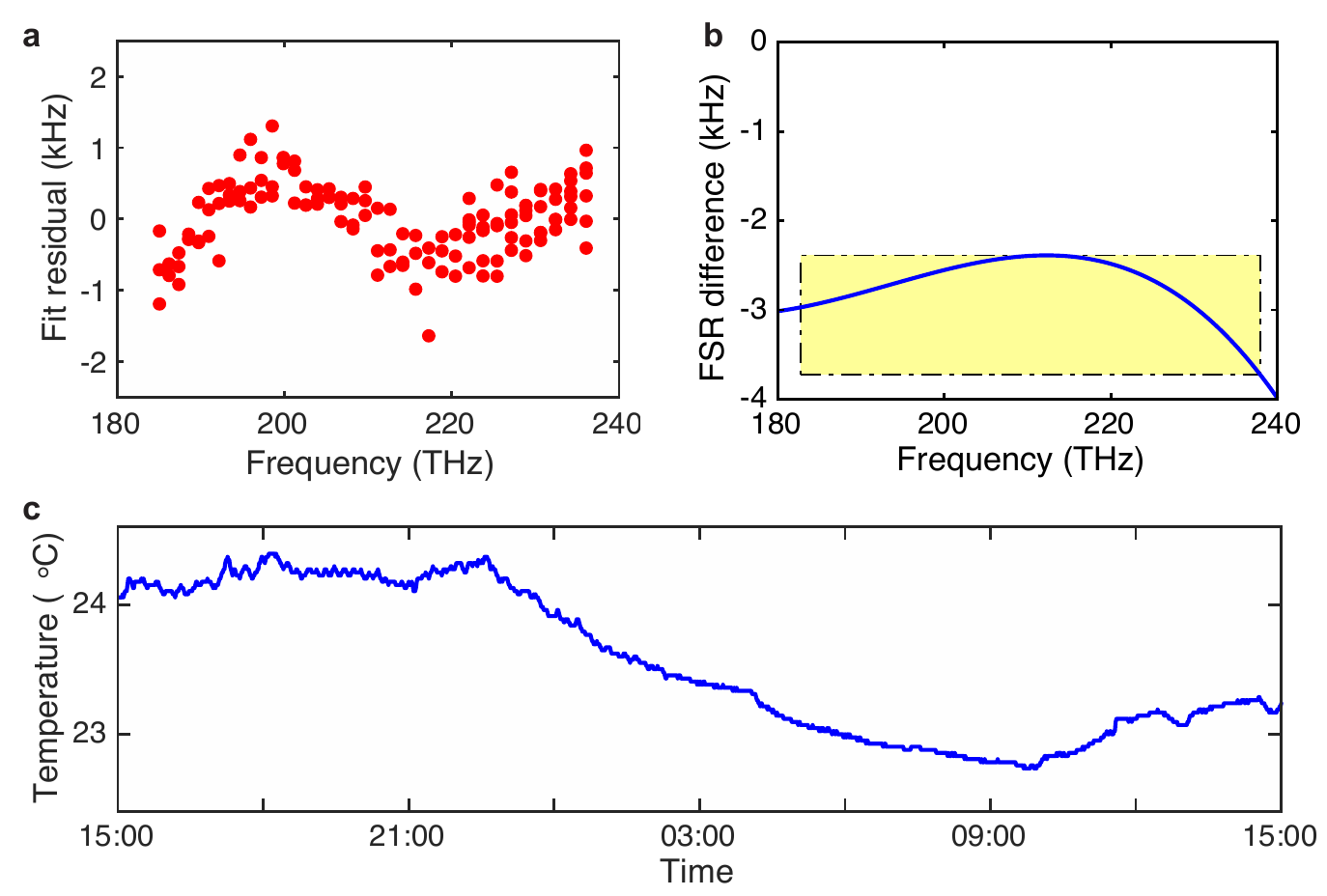 }
\caption{
\textbf{Characterization of the fiber cavity's temperature stability}.  
\textbf{a}. 
The fit residual of the fiber cavity's FSR using a quadratic polynomial fit, evidencing the necessity to use a cubic polynomial fit. 
\textbf{b}. 
The fiber cavity's FSR drifts with $\Delta T=9.3~^\circ$C temperature change. 
\textbf{c}. 
Measurement of the ambient temperature in our laboratory for 24 hours. 
}
\label{SIfig:residual}
\end{figure}

\vspace{0.8cm}
\section{Calibration and fit of the fiber cavity's dispersion}
\vspace{0.4cm}

In the main text, Fig. 2e shows the calibration of the fiber cavity's FSR $f_\text{fsr}$ that is frequency-dependent. 
This determines the accuracy of the relative frequency calibration in our VSA. 
We measure $f_\text{fsr}$ from 1260 to 1640 nm with an interval of 10 nm. 
The measured data are fitted with a cubic polynomial formula 
$f_\text{fsr}(\nu) = p_1\nu^3 + p_2\nu^2 + p_3\nu+p_4$, where $\nu$ is the optical frequency. 
The fit parameters are: $p_1 = 1.217(11)\times 10^{-7}$, $p_2 = -9.8(7)\times10^{-5}$, $p_3 = 0.026(2)$, $p_4 = 53.38(11)$. 

The reason why we use a cubic polynomial fit is the following.  
If we use a quadratic polynomial fit, the fit residuals (i.e. data deviations from the fit curve) are shown in Fig. \ref{SIfig:residual}a. 
The profile of residuals indicates that, due to our wide spectral measurement,  indeed a cubic polynomial fit (to the third order) is necessary. 


\vspace{0.8cm}
\section{Characterization of the fiber cavity's temperature stability}
\vspace{0.4cm}

We place the fiber cavity on a hot plate to investigate the influence of temperature drift. 
The fiber cavity's temperature is monitored by a thermistor thermometer with a measurement accuracy of $0.5~^\circ$C and a resolution of $0.01~^\circ$C. 
The temperature detector is placed on the coiled fiber cavity. 
Since the ambient temperature in our lab fluctuates within 1 degree over 24 hours, as shown in Fig. \ref{SIfig:residual}c, we do not apply any active temperature stabilization on our experimental setup. 

As shown in Fig. 2e in the main text, the two measured fiber dispersion curves at different temperatures are nearly identical except with a global relative shift in the y-axis. 
The FSR shift among the two curves versus frequency is shown in Fig. \ref{SIfig:residual}b. 
The overall FSR shift is around $-3$ kHz for $9.3^\circ$C temperature change, i.e. $-0.3$ kHz for $1^\circ$C change. 
Therefore, with this measured fiber cavity's FSR drift with temperature, once the ambient temperature is known, we can calculate the $f_\text{fsr}$ for a given temperature over the 55.1 THz frequency range. 

\begin{figure}[b!]
\centering
\renewcommand{\figurename}{Supplementary Figure}
\includegraphics{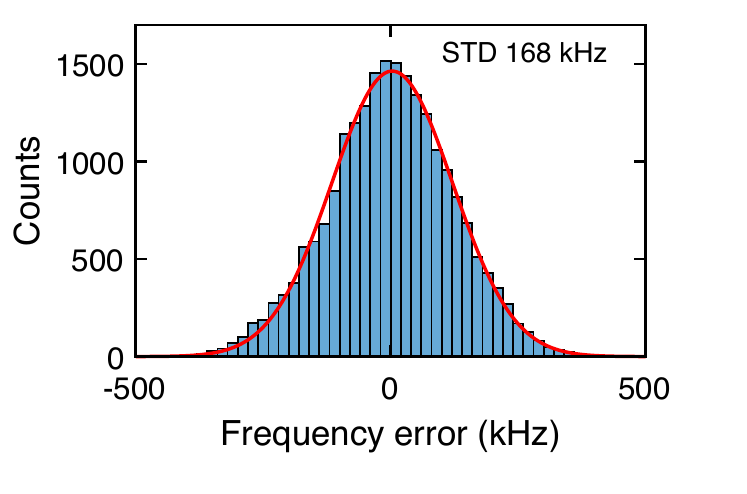}
\caption{
\textbf{Dead zone frequency error}. 
STD, standard deviation.}
\label{SIfig:Deadzone}
\end{figure}

\vspace{1cm}
\section{Accuracy and precision of the relative frequency calibration}
\vspace{0.5cm}
The fiber cavity only indicates the exact relative frequency at the resonances.  
In the experiment, to calibrate the relative frequency elsewhere, we simply assume the laser chirps linearly within the interval between resonances and utilize a linear interpolation. 
Thus it is necessary to evaluate the relative calibration accuracy and precision within the ``dead zone''. 

Here we adopt an optical frequency comb (OFC) assisted method \cite{Shi:23}. 
The chirp rate of the Santec laser is set to 50 nm/s.
A portion of the chirping laser is split out and interferes with the OFC.
The exact laser frequency can then be extracted from the beatnote signal using digital signal processing with a frequency tracking interval of 1 MHz.
To evaluate the frequency error introduced by linear interpolation in the ``dead zone'', we perform linear interpolation every megahertz with the sampled exact frequency at the fiber cavity's resonances. 
We then compare the OFC measured frequencies with the estimated value obtained by fiber cavity and linear interpolation, the error is statistically shown in Fig. \ref{SIfig:Deadzone} with a histogram. 
The center of the distribution is 3 kHz, evidencing the interpolation guarantees calibration accuracy and the standard deviation of the distribution reveals the calibration precision is 168 kHz. 

\vspace{1cm}
\section{Performance of the VSA}
\vspace{0.5cm}

\begin{figure}[b!]
\centering
\renewcommand{\figurename}{Supplementary Figure}
\includegraphics[width=0.95\columnwidth]{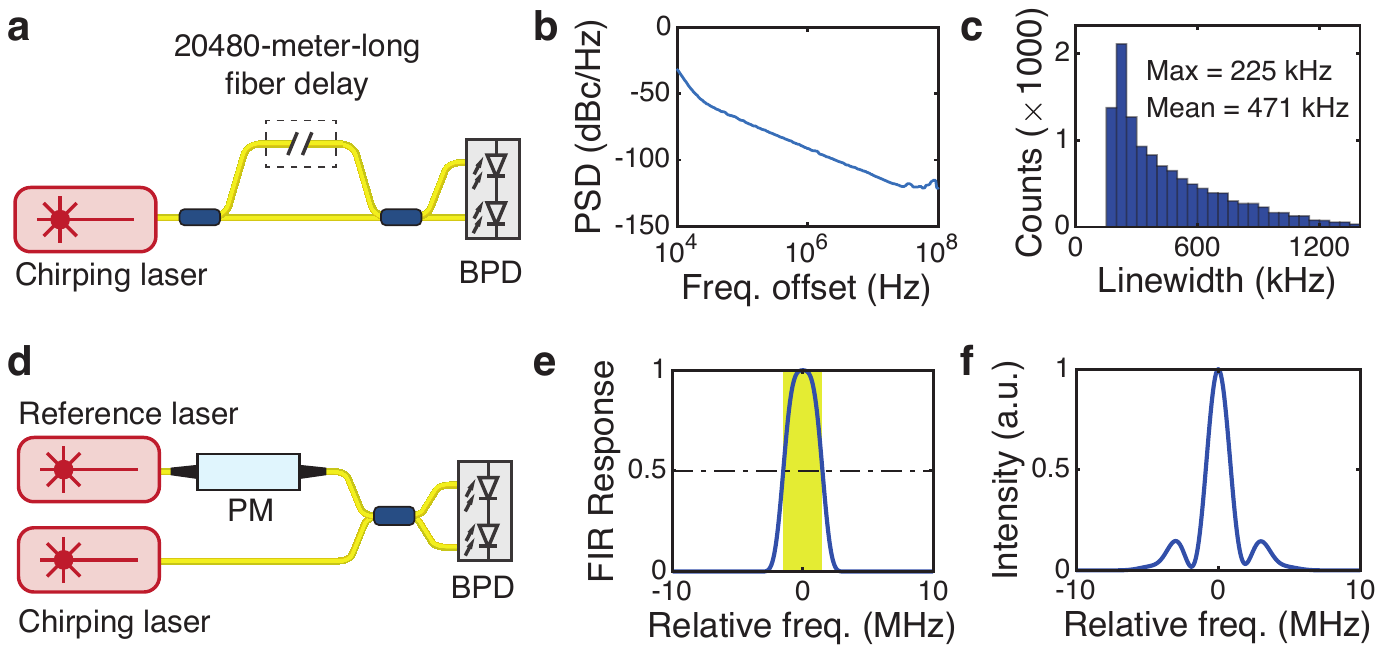}
\caption{
\textbf{Performance characterization of our VSA. a}. The self-delayed heterodyne measurement setup. 
\textbf{b}. Measurement of the static single-sideband frequency noise of the chirping laser.
\textbf{c}. The histogram of the chirping laser dynamic linewidth measured results, where 12207 100-$\mu$s segments of beatnote signals are taken and processed with FFT to extract the linewidth. 
\textbf{d}. The setup to verify the resolution bandwidth of spectrum dection of our VSA.
\textbf{e}. The frequency response of the FIR filter used in the VSA spectral measurement. 
\textbf{f}. The spectrum of phase modulated reference laser measured by our VSA, where two sidebands 3 MHz from the carrier can be ambiguously observed. 
}
\label{SIfig:Performance}
\end{figure}

In this section, we analyze the performance of the VSA in two aspects: the frequency resolution for passive devices characterizations, and the resolution bandwidth for the spectrum detection. 

In an ideal situation where the laser is perfect, the ultimate frequency resolution of the VSA is determined by the memory depth of the oscilloscope and laser chirp range. 
As stated in the main text, the ultimate frequency resolution is 99 kHz. 
However, if the laser linewidth is larger than 99 kHz, the data point corresponding to a specific frequency will mix with adjacent ones. 
The real frequency resolution is determined by the shortest slab of the system. 
Therefore it is critical to evaluate the laser's dynamic linewidth. 

The self-delayed heterodyne setup to measure the laser's dynamic linewidth is shown in Fig \ref{SIfig:Performance}a. 
The chirping laser is split into two branches. 
The upper branch passes through a 20408-meter-long fiber line to introduce a time delay $\Delta \tau$, which is longer than the coherent time of the laser.  
The upper branch is then combined with the undelayed lower branch. 
During the measurement, the laser chirps at 50 nm/s. 
The temporal beatnote signal is then directly recorded and transformed to the frequency domain with windowed fast Fourier transformation (FFT). 
The width of the FFT window $T_\mathrm{w}$ is inversely proportional to the FFT's frequency resolution $\delta f$ as $\delta f=1/T_\mathrm{w}$, which should be one order smaller than 99 kHz. 
In the experiment, we select $T_\mathrm{w} = 100$ $\mu$s, which corresponds to an FFT's frequency resolution $\delta f$ of 10 kHz. 
We measured 12207 100-$\mu$s windowed beatnote signals' linewidth, the histogram is shown in Fig. \ref{SIfig:Performance}c. 
The statistical laser linewidth corresponding to 100-$\mu$s is 471 kHz with a standard deviation of 292 kHz. 
Such a measured laser linewidth originates from multiple reasons, including the laser intrinsic linewidth, the laser chirp nonlinearity, and the instability of the fiber delay. 
It is hard to extract the linewidth induced by the laser intrinsic linewidth, thus we conclude 471 kHz as the frequency resolution of our VSA. 
Additionally, we also measure the frequency noise of the laser for reference. 
The laser is set to a specific frequency (i.e. 193.41449 THz) and combined with a fiber laser (NKT Koheras), the beatnote signal is directly analyzed with a phase noise analyzer (Rohde \& Schwarz FSWP)
The result is shown in Fig. \ref{SIfig:Performance}b. 

However, for spectrum detection, digital FIR filters of megahertz bandwidth are utilized. 
In this situation, to evaluate the VSA's performance, the resolution bandwidth is a more suitable parameter, which is the minimum spectral separation the spectrum analyzer can distinguish. 
In our experiment, the pass band 3-dB width of the FIR filter is designed to be 3 MHz, as shown in Fig. \ref{SIfig:Performance}e, thus we infer $3$ MHz as the spectrum measurement resolution bandwidth of our VSA. 
We further verify the result experimentally, the setup is shown in Fig. \ref{SIfig:Performance}d, where the chirping laser sweeps across the reference laser (NKT Koheras) with a chirp rate of 50 nm/s.
The reference laser is phase modulated with a 3 MHz sinusoidal signal to create sidebands. 
As shown in Fig. \ref{SIfig:Performance}f, the sidebands can be unambiguously distinguished. 

We finally compare the performance of our VSA with other commercial products and research advances, which is shown in Table \ref{tab:2}. Our VSA provides a new paradigm to characterize integrated nonlinear devices over an ultra-wide bandwidth with a globally high resolution. Meanwhile, the VSA also provides comparable spectrum analyzing ability as the most advanced spectrum analyzer. 

\begin{table*}
\caption{\label{tab:table2}
Performance comparison between multiple state-of-the-art OSAs, OVNAs, and our VSA. 
}
\begin{ruledtabular}
\begin{tabular}{c@{}ccccccc}
\multirow{2}*{Measure Mode} &  & YOKOGAWA & II$-$VI & EXFO & APEX & \multirow{2}*{OVNA\cite{Qing:19}} & \multirow{2}*{This work}\\
 &  & AQ6380 & 1500S & CTP10 & AP208xB &  & \\
\colrule\\[-6pt]
\multirow{2}*{Loss \& Phase} & Bandwidth & - & - & 52.9 THz & - & 1.075 THz & 55.1 THz \\
 & Resolution & - & - & 125 MHz & - & 334 Hz & 99 kHz \\[1pt]
\colrule\\[-6pt]
\multirow{3}*{Specturm} & Bandwidth & 68.2 THz & 5.3 THz & - & $\sim$15 THz & - & 55.1 THz\\
 & Resolution bandwidth & 625 MHz & 20 MHz & - & 5 MHz & - & 3 MHz\\ 
 & Dynamic range & 65 dB & 50 dB & - & 40 dB & - & 56 dB \\ 
\end{tabular}
\end{ruledtabular}
\label{tab:2}
\end{table*}

\vspace{1cm}
\section{OVNA phase measurement}
\vspace{0.5cm}

\begin{figure}[b!]
\centering
\renewcommand{\figurename}{Supplementary Figure}
\includegraphics[width=0.9\columnwidth]{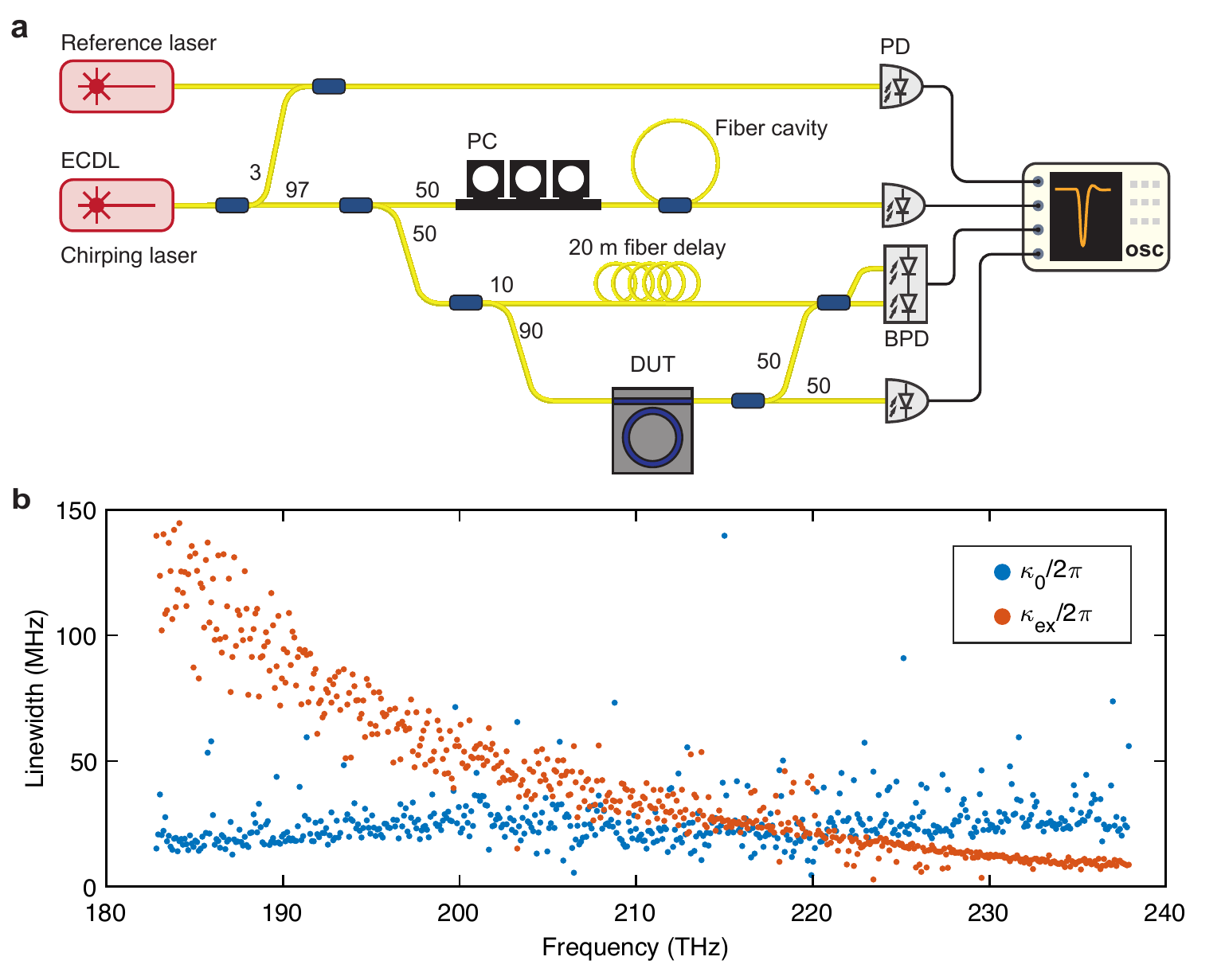}
\caption{
\textbf{Phase measurement}.  
\textbf{a}. Experimental setup.  
\textbf{b}. Measurement of resonance linewidth over the 55.1 THz spectral bandwidth.  
The phase measurement allows unambiguous determination of whether the resonance is over-coupled ($\kappa_\text{ex}>\kappa_0$) or under-coupled ($\kappa_\text{ex}<\kappa_0$).
}
\label{SIfig:OVNA}
\end{figure}

The setup of our VSA as an OVNA is shown in Fig. \ref{SIfig:OVNA}a. 
The branch containing a reference laser is for absolute frequency calibration. 
The branch containing the fiber cavity is for relative frequency calibration. 
Parallel to the branch where the laser transmits through the DUT, the laser passes through a 20-meter-long fiber, which introduces a time delay $\Delta\tau$. 
At the DUT's output, half of the laser is photodetected to generate the DUT's transmission trace.  
The other half interferes with the laser through the long fiber.  
In the following, we elaborate on how the phase information is extracted from the interference signal. 

If the laser chirps at the rate of $\gamma$, there is a frequency difference $\gamma\Delta\tau$ between the two branches due to the long fiber, and a beat signal is generated. 
Mathematically, the chirping laser from the long-fiber branch can be formulated as 
$$
    E_\text{upper} \propto \exp[i\omega_0(t-\Delta\tau) + i\gamma(t-\Delta\tau)^2/2]. 
$$
Similarly, for the DUT branch 
$$
E_\text{lower} \propto \exp[i\omega_0t + i\gamma t^2/2 + i\varphi(t)], 
$$
where $\varphi(t)$ is the phase shift introduced by the DUT. 
Therefore, the balanced photodetector output can be written as
$$
    I(t)\propto \mathrm{Re} (E_\text{upper}E_\text{lower}^*) = A\cos(\gamma\Delta\tau\cdot t + \varphi(t)), 
$$
where the constant phase $\omega_0\Delta\tau$ is merged into $\varphi(t)$, which can be canceled by introducing an additional reference path, and the high-order terms of $\Delta\tau$ are omitted. 
To extract the phase information $\varphi(t)$, Hilbert transformation is performed. 
In our case
$\mathcal{H}[I(t)] = -A\sin(\gamma\Delta\tau\cdot t + \varphi(t)), $
thus we have
\begin{equation}
\Phi(t) = \gamma\Delta\tau\cdot t+ \varphi(t) = -\arctan\frac{\mathcal{H}[I(t)]}{I(t)}. 
\label{eqn:OVNA}
\end{equation}
In our experiment, $\gamma\Delta \tau$ is obtained by linear fit of $\Phi(t)$ far from the resonances. 
The phase $\varphi(t)$ can be further extracted with $\Phi(t) - \gamma\Delta \tau\cdot t$. 
With the absolute and relative frequency calibration, $\varphi(t)$ is then transformed from the time domain to the optical frequency domain, and the phase response of the DUT $\varphi(f)$ is finally obtained. 

With our OVNA,  the microresonator's linewidth is characterized and shown in Fig. \ref{SIfig:OVNA}b.
The resonance frequency $\omega/2\pi$ and linewidth of each fundamental-mode resonance, ranging from $1260$ nm ($237.9$ THz) to $1640$ nm ($182.8$ THz), are measured.
For each resonance, the intrinsic loss $\kappa_0/2\pi$ and external coupling strength $\kappa_\text{ex}/2\pi$ are extracted by Lorentzian fit.
With our phase measurement of resonances, we can unambiguously determine whether the resonance is over-coupled ($\kappa_\text{ex}>\kappa_0$) or under-coupled ($\kappa_\text{ex}<\kappa_0$).

\vspace{1cm}
\section{Coherent LiDAR experiment}
\vspace{0.5cm}

\begin{figure}[t!]
\centering
\renewcommand{\figurename}{Supplementary Figure}
\includegraphics[width=0.9\columnwidth]{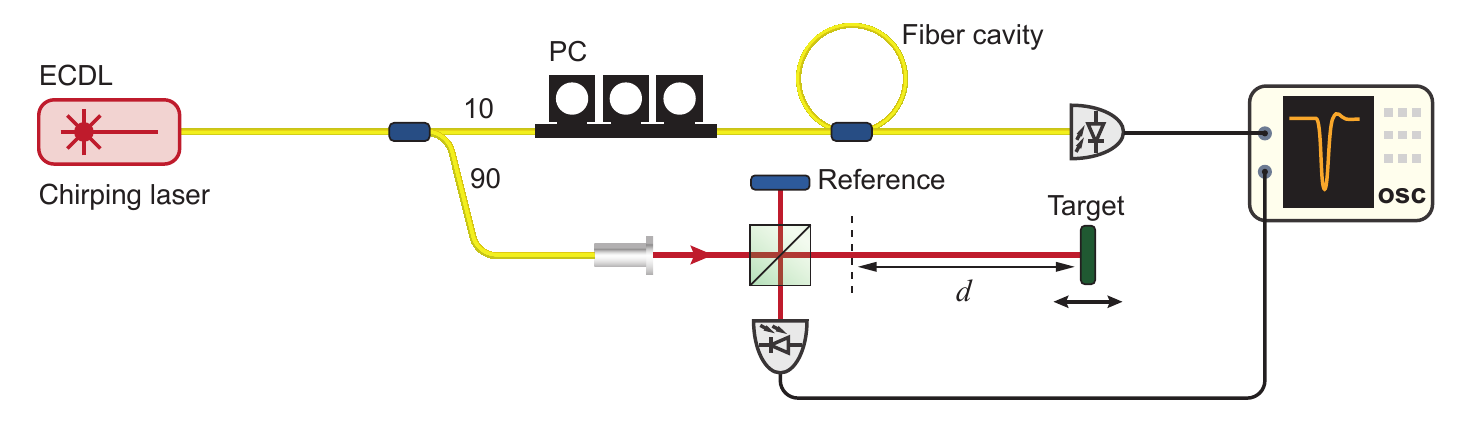}
\caption{
\textbf{Experimental setup of coherent LiDAR.} 
}
\label{SIfig:Ranging}
\end{figure}

\begin{figure}[t!]
\centering
\renewcommand{\figurename}{Supplementary Figure}
\includegraphics[width=0.6\columnwidth]{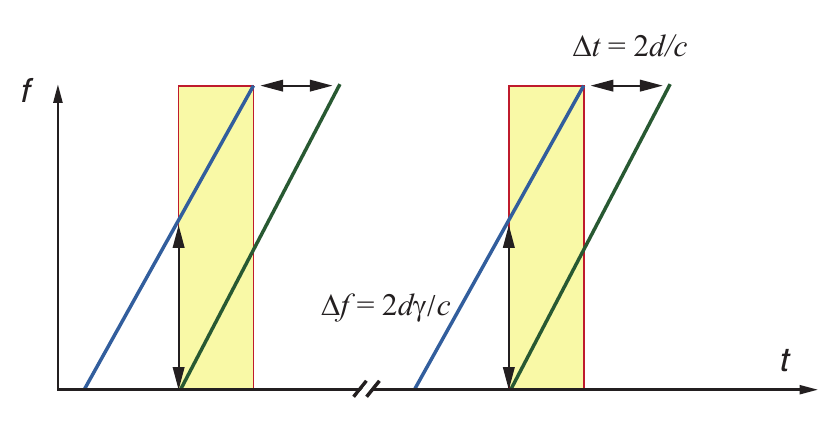}
\caption{
\textbf{Principle of coherent LiDAR.} 
}
\label{SIfig:Ranging_Principle}
\end{figure}

The experimental setup of coherent LiDAR is shown in Fig. \ref{SIfig:Ranging}. 
We use an ECDL that chirps from $192.2$ THz to $194.7$ THz.
The duration is $T=0.4$ s, thus the estimated linear chirp rate is $\gamma=6.25$ THz/s.
A 90:10 coupler is used to split the laser light into two parts. 
One part is directed to the VSA for frequency calibration, and the other is directed to a collimator and sent into free space. 

In the free space, the chirping laser is split into two paths with path difference $d$ and then recombined to interfere on a photodetector (FEMTO, HBPR-450M).
As shown in Fig. \ref{SIfig:Ranging_Principle}, the path difference $d$ introduces a time delay $\Delta t=2d/c$,
which generates a beat signal with frequency of $\Delta f=\gamma(t) \Delta t$, where $\gamma(t)$ is the time-dependent chirp rate. 
Thus, the beat signal can be written as:
\begin{equation}
\begin{aligned}
V(t)&\propto\cos(2\pi\Delta f t)\\
&=\cos\left(2\pi\frac{2d}{c}\gamma(t) t\right)
\end{aligned}
\label{Eq.Vt}
\end{equation}
From Eq. \ref{Eq.Vt} we can see that the beat signal's frequency is $\Delta f=2d\cdot\gamma(t)/c$, which can be obtained with fast Fourier transformation (FFT).  
In the ideal case where the chirp rate is constant, the path difference $d$ can be directly extracted from the obtained $\Delta f$. 
In reality, the laser does not chirp linearly, i.e. the chirp rate $\gamma(t)$ fluctuates. 
Thus to extract $d$ from $\Delta f$, an accurate trace of $\gamma(t)$ is required. 

Fortunately, our VSA allows the monitor and record of the instantaneous laser frequency during its chirping.
With known instantaneous laser frequency, the chirp rate $\gamma (t)$ can be calculated.
Thus with calibrated $\gamma (t)$ we can re-scale the signal's time axis by $t'=\gamma(t) t$ to
\begin{equation}
V(t')=\cos\left(2\pi\frac{2d}{c} t'\right).
\label{Eq.Vt'}
\end{equation}
Then the precise range profile can be retrieved. 

In addition to chirp rate calibration, zero-padding \cite{Shinpaugh:92} is also implemented in data processing, which inserts zeros before and after the measured data.
As mentioned in the main text, the ranging resolution $\delta d$ is limited by chirp bandwidth $B$ as $\delta d=c/2B$ due to the constraints of the FFT.
Zero-padding increases the bandwidth, which is useful for resolving ambiguities and reducing the quantization error in estimating the spectral peaks \cite{Shinpaugh:92}. 
Despite this, zero padding can introduce additional noise through spectral leakage \cite{Zhang:01}.
Given our broad measurement bandwidth and high precision of the relative frequency calibration, the beat signal exhibits a high signal-to-noise ratio (SNR). This high SNR enables accurate peak detection using 8192-fold zero-padding.
It should be noted that zero-padding cannot improve the ranging resolution, i.e. resolving two objects distant less than $\delta d=c/2B$.

\clearpage
\section*{Supplementary References}
%